# The transient route to resonance

Henning U. Voss

Cornell University, Ithaca, NY, USA

**Abstract —**The periodically driven harmonic oscillator with damping is one of the most elementary and trusted models in physics and normally applied in its steady state, disregarding specific initial conditions and associated transients. For example, steady state solutions are already sufficient to describe resonance. In some applications transient solutions might play a role though, and a review of these might be useful. In order to characterize the transients of the harmonic oscillator, its winding number is graphed in dependence of initial conditions. These visualizations show a complexity of the transients of the harmonic oscillator that are not readily evident from the inspection of the model alone.

**Key words—Harmonic oscillator, Resonance, Damping, Arnold tongues, Devil's staircase**

A. Introduction

The periodically driven harmonic oscillator with damping serves as a trusted model from the description of elementary particle interactions to technical applications such as radio signal transmission and magnetic resonance imaging. In most applications, the harmonic oscillator is only considered in its steady state, by disregarding the dependence of its transient solutions from specific initial conditions. Disregarding transients usually does not take away any accuracy of this model, as for most applications transients die reasonably fast and are not of any practical concern. For example, the steady state solution of the periodically driven harmonic oscillator with damping completely describes the characteristic resonance curve that permeates almost all areas of physics.

However, there are applications of resonance phenomena in which it might be beneficial to consider transients. For example, the human heart acts as a driving force to the vascular system, which can be modeled as a network of RLC circuits, potentially exhibiting transient and resonance effects. It could be generally useful to understand the effects of transients on the route to resonance in these evolutionary designed, rather than man-made, systems. In addition, transient oscillations are responsible for the characteristic sound of many musical instruments.

The governing equations of resonance and transients in the harmonic oscillator are well known, and there should be no surprises. However, when categorizing transients by their winding number, or the number of amplitude reversals of the oscillator divided by the number of amplitude reversals of the driving force, complex patterns arise that might not be readily evident by inspection of the analytical solutions. In particular, there is a significant dependence of the winding number on the three initial conditions of the system: The amplitude, the change of amplitude, and the driving phase. In addition, winding numbers of the undamped oscillator depend on the driving frequency in a discontinuous fashion, with sharp changes for values of the driving frequency that correspond to small integer fractions of the resonance frequency of the oscillator. The corresponding damped oscillator parameter space shows tongue-shaped areas with their tips at these singled-out frequencies, defining frequency locking regions.

This transient route to resonance is investigated in the following. First, the resonance curve for the harmonic oscillator is reviewed, then analytic solutions for the undamped harmonic oscillator are derived, and their dependence on drive frequency and initial conditions is being visualized. Finally, the effects of damping on this system is investigated numerically.

## B.  Model and resonance

The periodically driven damped harmonic oscillator model is given as a non-autonomous ordinary differential equation for the time-dependent oscillator amplitude y(t),

$$\ddot{y}(t) + \omega_0^2\, y(t) + \beta_0\, \dot{y}(t) = D\, x(t)\,, \qquad \text{with } x(t) = \cos(\omega t + \Delta)\,. \tag{1}$$

The parameter $\omega_0 > 0$ is the angular resonance frequency that would occur without damping, D > 0 is the drive amplitude, $\omega > 0$ the angular drive frequency, and $\Delta$ the drive phase offset. The damping constant $\beta_0$ is assumed to be larger or equal to zero, but smaller than the critical damping $\beta_{\text{crit}} = 2\omega_0$. For the underdamped oscillator with damping much smaller than critical damping, which is the only case considered here, $\omega_0$ is close but not identical to the resonance frequency with damping,

$$\omega_d = \sqrt{\omega_0^2 - \frac{\beta_0^2}{4}}\,. \tag{2}$$

In case of a mass-spring system with damping, y(t) would be the position and $\dot{y}(t)$ the velocity of the mass. The parameters would be $\beta_0 = \frac{c}{m}$, where c is a friction constant, m the mass, and $\omega_0 = \sqrt{\frac{k}{m}}$, with k the spring constant. In case of an electric RLC circuit as being used for example in MRI signal reception , y(t) would be the charge in the capacitor and $\dot{y}(t)$ the current. The other parameters would be $\beta_0 = \frac{R}{L}$, with R the resistance and L the inductance, and $\omega_0 = \frac{1}{\sqrt{LC}}$, with C the capacitance. The value $\beta = \frac{\beta_0}{2}$ is usually called the attenuation of the RLC circuit. The parameter D would be defined by the coupling to a received radiofrequency signal with amplitude x(t). However, all the following considerations are independent of the specific physical interpretation of Eq. (1).

In order to derive the resonance curve for the system described by Eq. (1), the harmonic oscillator is being described as an input/output system, where x(t) is the input and y(t) the output. Since the dynamics of y(t) is linear in x(t), its stationary or steady state behavior can be analyzed by the frequency response function of the oscillator. With f being frequency, $\omega = 2\pi f$, $x(t) = \int X(\omega)e^{i\omega t}d\omega$, and $y(t) = \int Y(\omega)e^{i\omega t}d\omega$, the input/output relationship between the Fourier transforms of x(t) and y(t) is given by

$$Y(\omega) = H(\omega)X(\omega)\,,$$

in which

$$H(\omega) = \frac{D}{\omega_0^2 - \omega^2 + i\beta_0\omega}$$

is the frequency response function of the oscillator. If expressed in amplitude and phase notation, $H(\omega) = G(\omega)e^{i\Phi(\omega)}$, amplitude and phase are given by

$$G(\omega) = \frac{D}{\sqrt{(\omega_0^2 - \omega^2)^2 + (\beta_0\omega)^2}} \quad \text{and} \quad \Phi(\omega) = \arg(\omega_0^2 - \omega^2 - i\beta_0\omega). \tag{3}$$

The amplitude $G(\omega)$ is of particular importance as it determines the amplitude of the oscillator in response to the drive amplitude in the time domain, too. It defines the resonance curve shown in Figure 1. This curve is of even more general applicability than described in Eq. (1), as the input does not need to be a cosine function but can be in fact any signal with a properly defined Fourier transform.

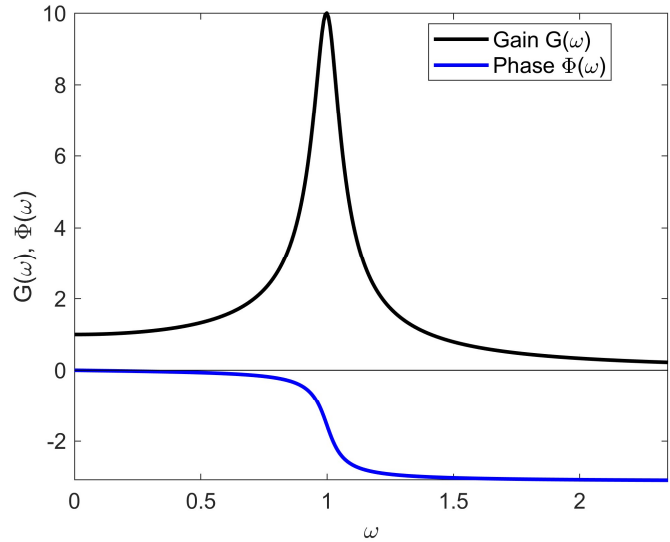

**Figure 1 (color): The resonance curve of the driven harmonic oscillator with damping.**
Shown are gain and phase of the transfer function $H(\omega)$ with the following parameters (used throughout this article if not mentioned otherwise): $\omega_0 = D = 1$, $\beta_0 = 0.1$. At resonance, these parameters yield a maximum gain of 10 at the resonance frequency, as follows from Eq. (3). All calculations made with MATLAB R2020b (The MathWorks, Inc., Natick, MA).

## C. General solution for the case without damping

The general solution of Eq. (1) is the superposition of the special solution of the driven oscillator and the general solution of the oscillator without drive. Equivalently, it is the sum of the steady state oscillation and transient solutions. For the case without damping ($\beta_0 = 0$), the general solution can be written as

$$y(t) = Y_{ss}e^{i(\omega t + \Delta)} + Y_{tr}e^{i(\omega_0 t + \Phi)}. \tag{4}$$

All signals are to be understood as being the real part of complex number expressions without further indication. Inserting $y(t) = Y_{ss}e^{i(\omega t + \Delta)}$ into the driven oscillator equation gives

$$Y_{ss} = \frac{D}{\omega_0^2 - \omega^2} \ . \tag{5}$$

This is the same amplitude that had been derived before in the frequency domain, which is plausible as this part of the solution defines the steady state, an assumption in that derivation.

The remaining unknowns $Y_{tr}$ and $\Phi$ depend on the triplet of initial conditions $(y_0, v_0, \Delta)$. The coefficients $y_0$ and $v_0$ are the initial amplitude and change of amplitude, and $\Delta$ is the initial value for the phase of the drive signal. Note that in most textbooks about the harmonic oscillator the latter initial condition is neglected, but here it is retained because it adds information. From

$$y_0 = y(0) = Y_{ss}e^{i\Delta} + Y_{tr}e^{i\Phi}, \qquad v_0 = \dot{y}(0) = i\omega Y_{ss}e^{i\Delta} + i\omega_0 Y_{tr}e^{i\Phi}$$

and the fact that the signals defined by both quantities should be real, follows

$$y_0 = Y_{ss}\cos(\Delta) + Y_{tr}\cos(\Phi), \qquad v_0 = -\omega Y_{ss}\sin(\Delta) - \omega_0 Y_{tr}\sin(\Phi) \ .$$

From the expression for the cosine of an arctan function, the unknowns $Y_{tr}$ and $\Phi$ can now be computed. With

$$\alpha_1 = \frac{v_0 + \omega Y_{ss}\sin(\Delta)}{\omega_0}, \qquad \alpha_2 = Y_{ss}\cos(\Delta) - y_0$$

follows

$$Y_{tr} = -\text{sign}(\alpha_2)\sqrt{\alpha_1^2 + \alpha_2^2}, \qquad \Phi = \arctan\left(\frac{\alpha_1}{\alpha_2}\right) \ . \tag{6}$$

Equations (4) to (6) constitute the analytic solution for the undamped driven harmonic oscillator in dependence of the initial conditions for its amplitude, change of amplitude, and drive phase. They form the basis for the visual analysis to follow.

(There are some isolated parameter triplets for which these expressions are not clearly defined, for example $(y_0, v_0, \Delta) = (0, 0, \pi/2)$, and these expressions would have to be modified accordingly. Also, the exact resonance of the undamped oscillator [1] is not further considered here.)

## D. Undamped solutions in dependence of initial conditions

The following visualizations are performed with a time step of 0.01 and $\beta_0 = 0$. The initial change of amplitude is $v_0 = 0$. In case of a mass-spring system, this would correspond to the natural situation of holding the mass at position $y_0$ and then letting it go, and switching on the driving force, at time t = 0.

A typical solution of Eq. (1) is displayed in Figure 2A. The number of drive periods is 10, shown as thin lines in the graph. The solution for y(t), Eq. (4), is more involved, as it is the linear superposition of the two oscillatory terms in Eq. (2), each with its own amplitude, frequency, and phase offsets. The underlying number of periods of the superposition signal can be estimated by counting the extrema of the signal. In this case, there are 67. Denoting the number of extrema of the harmonic oscillator with m and the number of extrema of the drive with n, this defines a *winding number* of m:n. In this example,

m:n = 67/20 = 3.35. This winding number will be the main parameter used for the visualization of the transient solutions of the driven harmonic oscillator. Practically, the number of extrema is computed as the number of sign changes of the analytical solutions' time derivative.

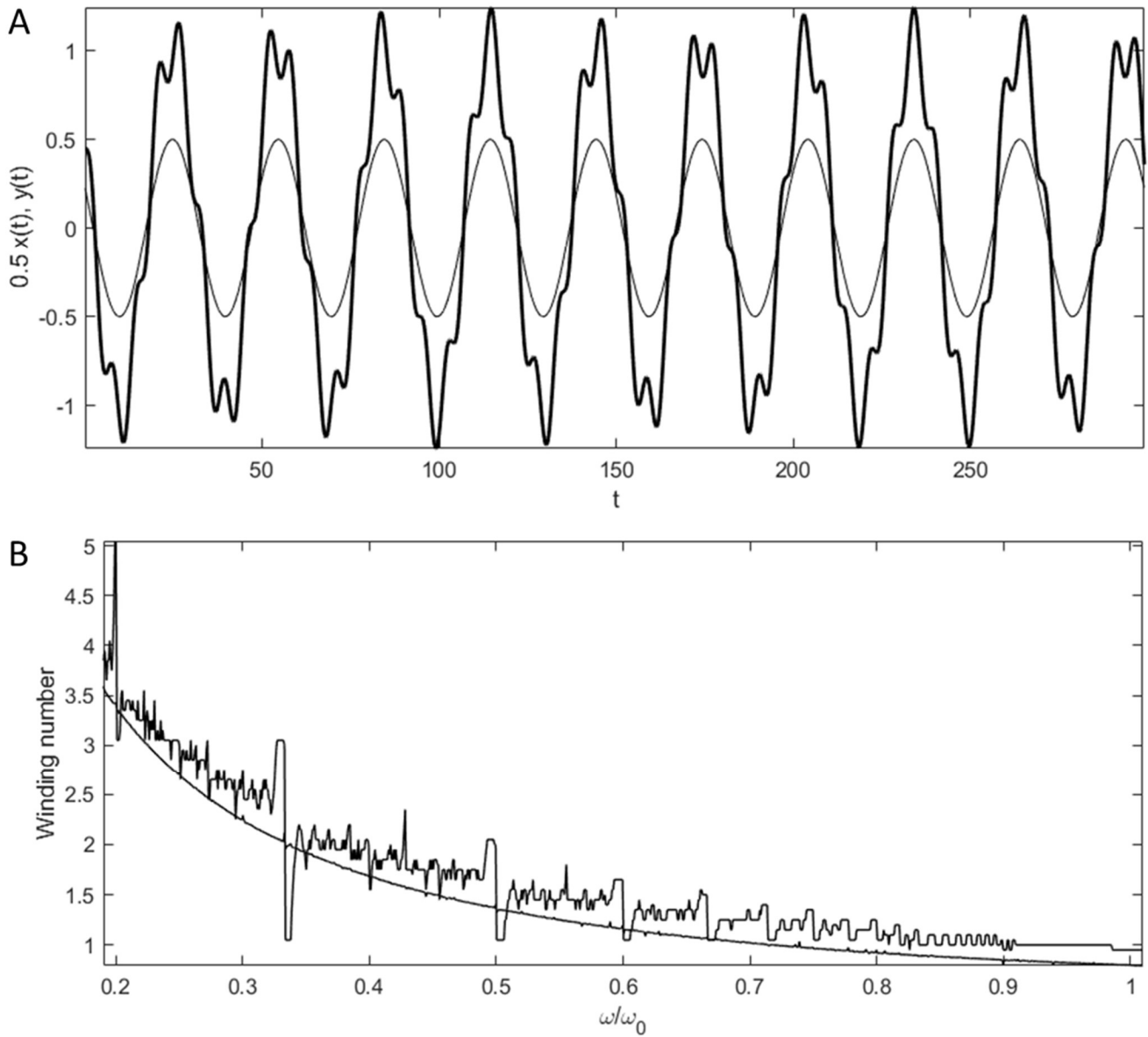

**Figure 2: Solution and winding numbers for the driven harmonic oscillator without damping.**
A) One specific solution for initial condition $(y_0, v_0, \Delta) = (0.45, 0, 1.1)$, at drive frequency $\omega = 0.21\,\omega_0$. While the drive (thin line) has 20 extrema, the solution has 67 extrema, defining a winding number of 67/20 = 3.35.
B) The winding number in dependence of angular drive frequency fluctuates for some parameters. Two solutions for the same initial conditions as in panel A are shown, one with 10 (ragged graph) and one with 1000 drive periods (smoother graph; offset by -0.2 for clarity).

A variation of the drive frequency ratio $\frac{\omega}{\omega_0}$ over a range from about 0.2 to 1, which corresponds to the resonance frequency, provides more insight into the overall behavior of the driven harmonic oscillator on its transient route to resonance. It is shown in Fig. 2B. For a small number of drive periods, the winding number in dependence of drive frequency fluctuates considerably, and these fluctuations reduce when the number of drive periods increases. Therefore, this is a finite sample effect on the infinitely lasting undamped transients.

The question arises if this behavior of the harmonic oscillator is generic or exists for certain isolated parameters only. In order to answer this question, the *aggregate winding number* over all tested drive frequencies is defined next as the mean of the absolute value of the discrete derivative of the winding number. This quantity eliminates the trend in the winding number and captures the fluctuation differences shown in Fig. 2. For example, the ragged graph has an aggregate of 0.050, whereas the smoother graph has an aggregate of only 0.006. Figure 3 displays the aggregate in the ($y_0$, $\Delta$) parameter space, where $v_0 = 0$ is being kept constant. As expected, the aggregate has a point symmetry (Figure 3 inset), and only one quadrant of the plane is shown in the main panel, therefore. To answer the question: Complex behavior of the winding number is generic and does not occur only for isolated situations. It fills a large part of the parameter space (non-blue regions in Fig. 3). In addition, the aggregate landscape appears to be quite detailed for the simple system of a driven harmonic oscillator without damping. Of note is that for the usually considered case of Eq. (1) with $\Delta = 0$, or a plane cosine drive, there is only a small range of amplitude initial conditions around $y_0 = 0.8$ with interesting behavior.

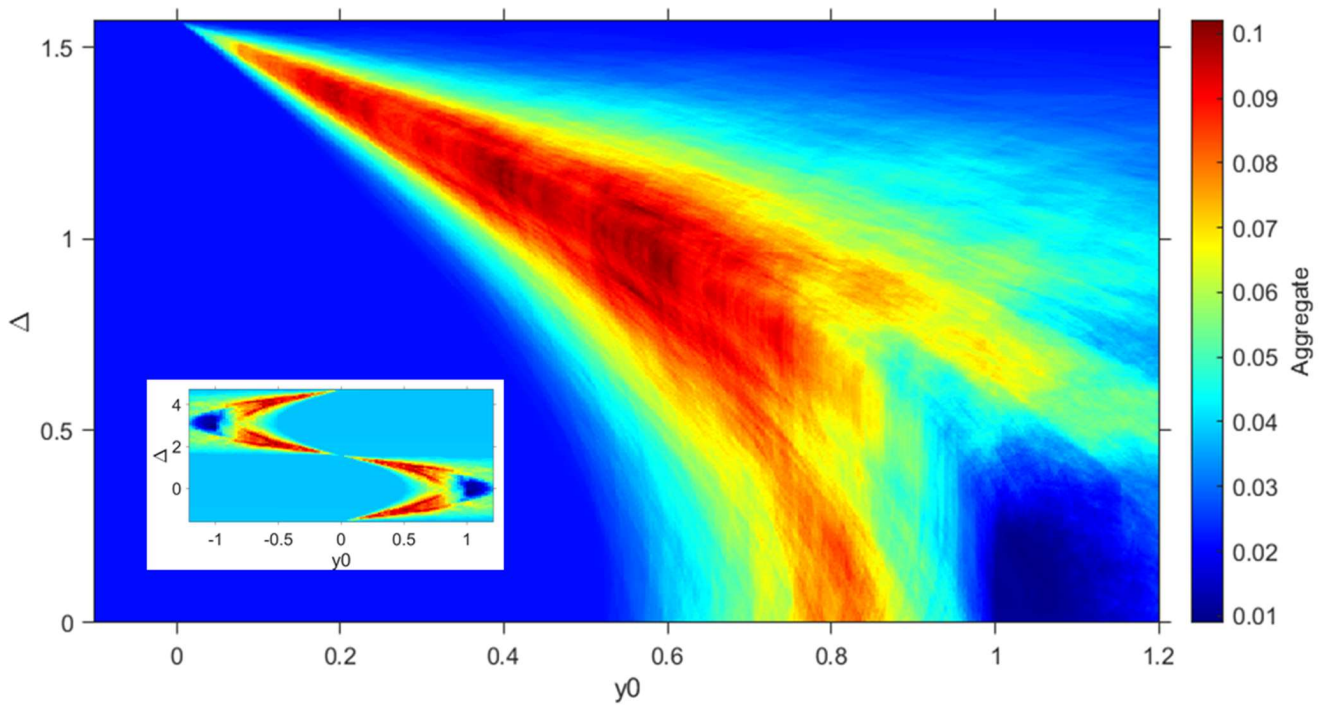

**Figure 3 (color): Parameter space for the driven harmonic oscillator without damping.**
The parameter space is defined by $y_0$ (abscissa) and $\Delta$ (ordinate), and color encodes the aggregate winding number, defined over a range of drive frequencies. The resulting color map is quite detailed for such a simple linear system but can be fully described by the general solution of the driven harmonic oscillator Eq. (2). The inset displays the symmetry of the parameter space, with its center at ($y_0$, $\Delta$) = (0, $\pi/2$). (The initial change of amplitude is $v_0 = 0$. The number of oscillations per point in the parameter space is 20 and $\omega/\omega_0$ is varied over the same range as in the previous examples.)

## E. Damped solutions in dependence of initial conditions

In contrast to all previous results, Eq. (1) is now solved numerically with a solver for ordinary differential equations to account for damping, which is not part of the analytic solution Eq. (4). The resonance frequency now decreases to the value in Eq. (2).

Figure 4A shows that the parameter space for the driven harmonic oscillator with damping is dominated by tongue-shaped regions of winding number m:n = 1:1 with the tip of the tongue pointing to small integer ratios of the drive frequencies. If displayed with a different color scale (Fig. 4C), a fractal or self-similar appearance of the damped, driven harmonic oscillator becomes evident.

The tongues in parameter space superficially resemble Arnold tongues as they are observed in nonlinear maps (only) [2]. There is a crucial difference between the observed tongues and Arnold tongues: The frequency locking ratio inside the tongues is always 1:1 and does not correspond to the winding number m:n at the tip of the tongues. Therefore, a cut through the parameter plane defined by a constant $\beta_0 > 0$ does not yield a Devil's staircase with plateaus at increasing values for m:n [3] but locally plateaus at m:n = 1:1 (Fig. 4B).

## F. Discussion and conclusion

I have tried to visualize how the solutions of the driven harmonic oscillator, one of the most elementary models in physics, depend in an intricate way on their initial conditions. This is true for both the undamped oscillator, for which the analytic solutions were visualized, as well as the damped oscillator, for which numerical solutions were considered. This exercise was inspired by the article "Mechanical resonance: 300 years from discovery to the full understanding of its importance" [1]. It mentioned that "the weakly damped oscillator in its transient state can show motions of more or less complicated shapes which certainly are not intuitively perceived as a direct consequence of the initial conditions" and that in systems without damping, unless the driving and resonance frequencies "are in proportion of small natural numbers, … this motion can appear quite irregular, depending on the initial conditions, and remains so for all times." The approach to shed more light onto these properties of transients consisted of two crucial ingredients: The winding number, a concept almost exclusively used otherwise in nonlinear dynamics, for studying the finite time and transient solutions of the harmonic oscillator, and the inclusion of the third initial condition, the phase of the drive.

It is the hope of the author that the visualization of transients in the route to resonance inspires the deeper investigation of the dynamics of physiological systems such as the cardiovascular circulation. The importance of resonance as a mechanism for pulse pressure amplification [4, 5] and the irregularity of the healthy heart beat could provide a dynamic setting close to a permanently transient state near resonance. This would have implications for the flow dynamics of the cerebrospinal fluid of the brain, an active research topic for its importance in neurodegenerative disease [6]. Novel ultrafast imaging methods that visualize the pulsatile properties of the human brain could shed light onto these phenomena [7, 8]. Notably, m:n phase locking and even Arnold tongues have been observed in the human cardiorespiratory system [9, 10] and musical instruments [11]. Whether the here observed tongues in the parameter space can be seen as a general precursor phenomenon of driven nonlinear or synchronizing nonlinear oscillators would be an interesting topic for further investigation.

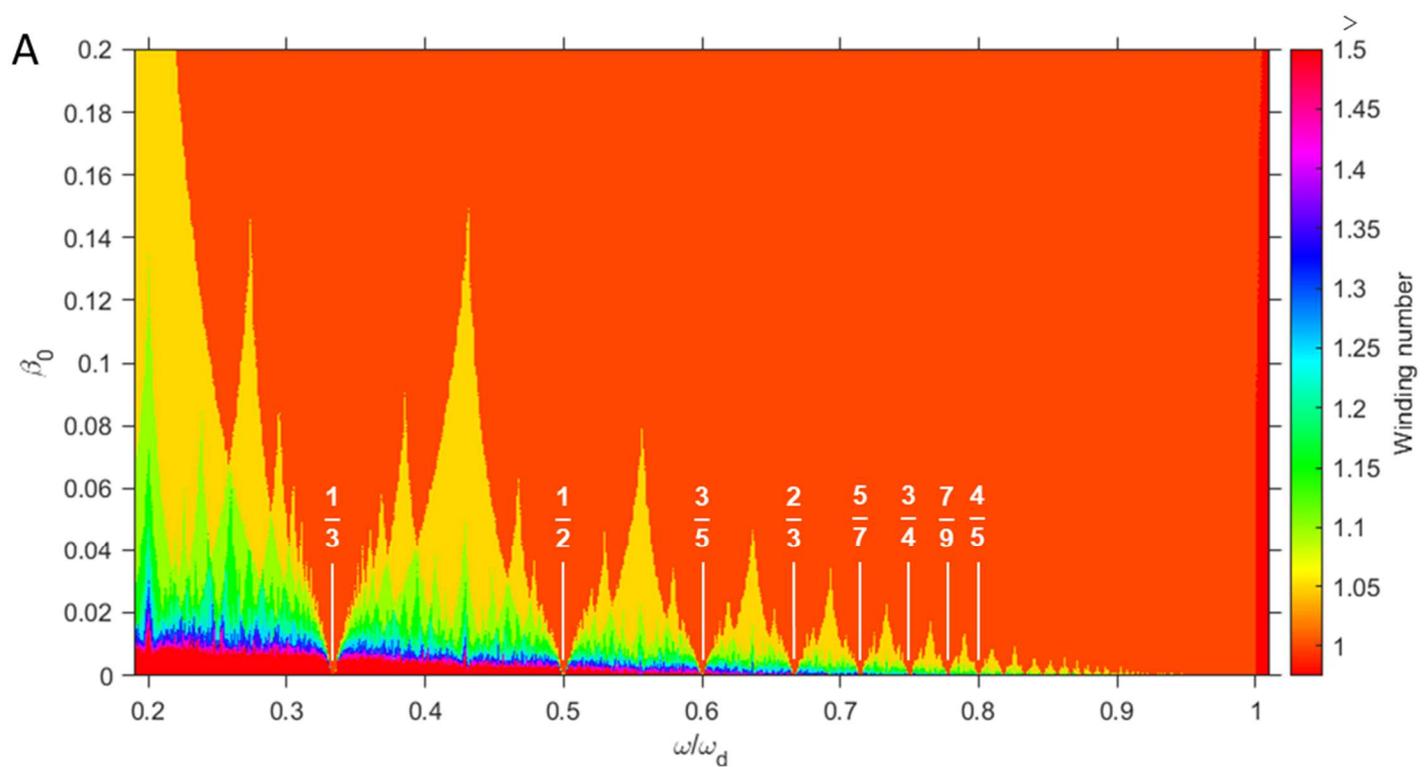

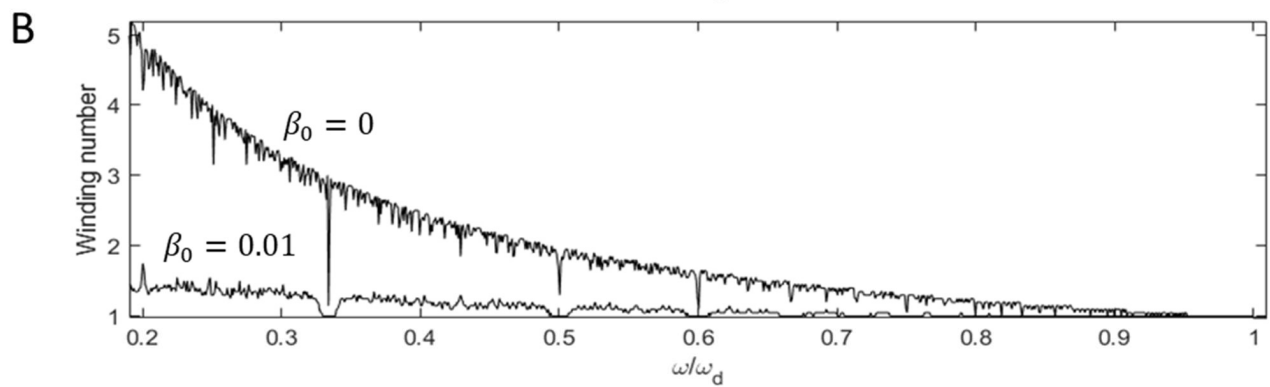

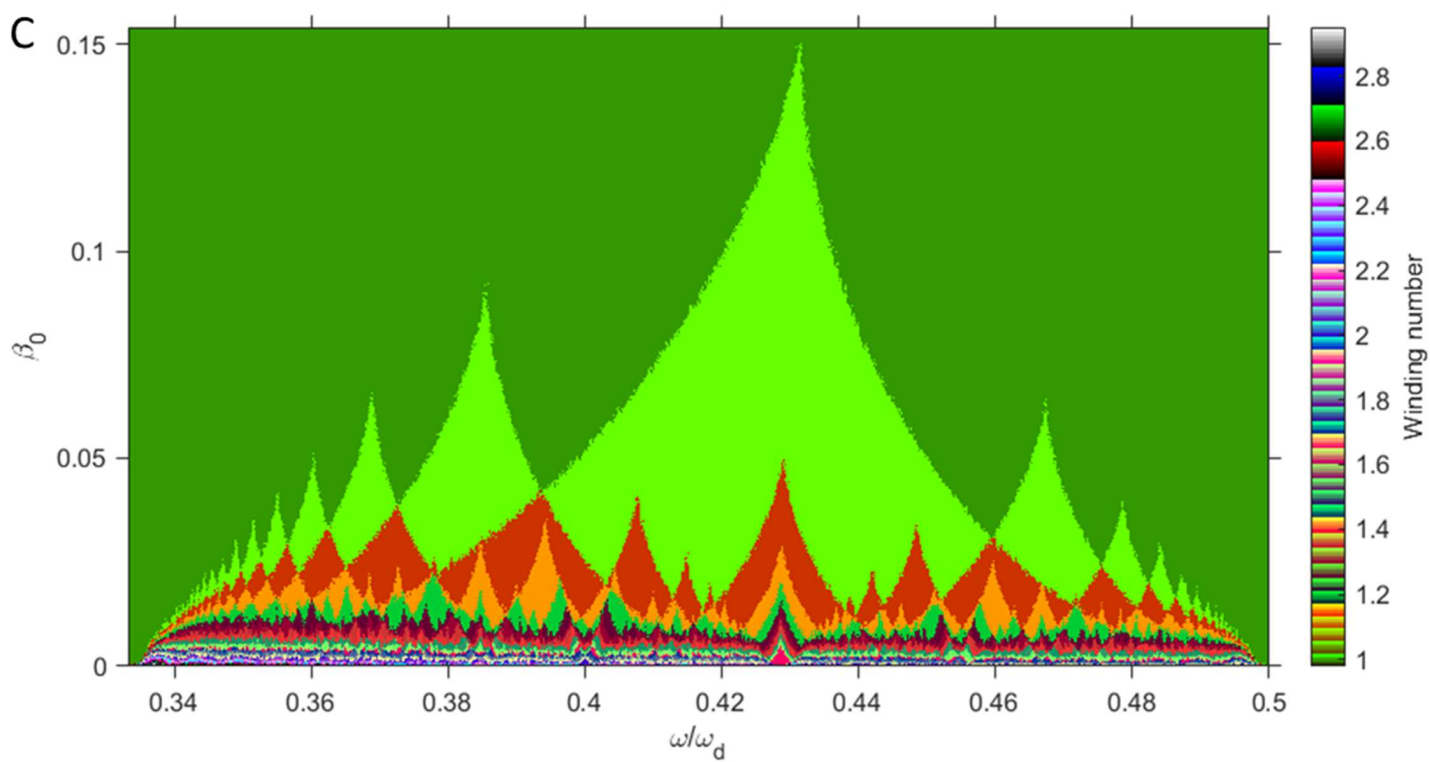

**Figure 4 (color): Parameter space for the driven harmonic oscillator with damping.**
A) The parameter space is dominated by tongue-shaped regions of winding number = 1:1 (orange) with the tongue tips pointing to small integer ratios of the drive frequencies.
B) The winding number for the case without damping, which corresponds to the $\beta_0 = 0$ line in panel A, and with some damping, $\beta_0 = 0.01$.
C) Detail of the parameter space between the 1/3 and 1/2 tongue tips of panel A to enhance the fractal-like appearance of the driven harmonic oscillator.
In all panels, the number of simulated oscillations per point in the parameter space are 20, and $(y_0, v_0, \Delta) = (0.001, -0.001, \pi/2)$.

_________________________________________